\documentclass[amsmath,aps,twocolumn,prl,showpacs,showkeys,groupedaddress]{revtex4}
\usepackage{cases}
\usepackage{dsfont}
\usepackage{xspace}
\usepackage{comment}
\usepackage{amsmath}
\usepackage{amssymb}
\usepackage{filecontents}
\usepackage{graphicx}

\DeclareMathAlphabet{\matheub}{U}{eur}{m}{n}

\newcommand{\dint}{\textrm{d}}

\newcommand{\OC}{\mathcal{O}}

\newcommand{\RealPart}[1]{\mathchoice{\Re\left(#1\right)}{\Re(#1)}{\Re(#1)}{\Re(#1)}}

\newcommand{\latin}[1]{{\it #1}}
\newcommand{\ie}{\latin{i.e.}\@\xspace}

\newcommand{\half}{\mathchoice{\frac{1}{2}}{(1/2)}{\frac{1}{2}}{(1/2)}}
\newcommand{\Exp}[1]{\operatorname{exp}\left(#1\right)}
\renewcommand{\exp}[1]{\mathchoice{e^{#1}}{\operatorname{exp}\left(#1\right)}{\operatorname{exp}\left(#1\right)}{\operatorname{exp}\left(#1\right)}}

\newcommand{\elabel}[1]{\label{eq:#1}}
\newcommand{\eref}[1]{(\ref{eq:#1})}
\newcommand{\Eref}[1]{Eq.~(\ref{eq:#1})}

\newcommand{\flabel}[1]{\label{fig:#1}}
\newcommand{\fref}[1]{Fig.~\ref{fig:#1}}

\usepackage{dsfont}
\newcommand{\gpset}[1]{\mathds{#1}}
\newcommand{\Rset}{\gpset{R}}
\newcommand{\Zset}{\gpset{Z}}

\newcommand{\bra}[1]{\left\langle#1\right|}
\newcommand{\braket}[2]{\left\langle#1|#2\right\rangle}
\newcommand{\ket}[1]{\left|#1\right\rangle}

\newcommand{\anglePeriod}{\xi}
\newcommand{\timePeriod}{\psi}
\newcommand{\syncPeriod}{S}
\newcommand{\syncTime}{\tau}

\newcommand{\deltat}{\delta t\,}
\newcommand{\Jtilde}{\tilde{J}}
\newcommand{\thetatilde}{\tilde{\theta}}
\newcommand{\thetabar}{\bar{\theta}}
\newenvironment{subeqnarray}[1]{\begin{subequations}#1\begin{eqnarray}}{\end{eqnarray}\end{subequations}}

\newcommand{\SectionHead}[1]{{\it #1 -- }}

\begin{document}
\title{Doppler synchronization of pulsating phases by time delay}
\author{Gunnar Pruessner}
\email{g.pruessner@imperial.ac.uk}
\affiliation{Department of Mathematics,
Imperial College London,
180 Queen's Gate,
London SW7 2BZ, United Kingdom}
\author{Seng Cheang}
\affiliation{Department of Mathematics,
Imperial College London,
180 Queen's Gate,
London SW7 2BZ, United Kingdom}
\author{Henrik Jeldtoft Jensen}
\email{h.jensen@imperial.ac.uk.}
\affiliation{Department of Mathematics and Complexity \& Networks Group,
Imperial College London,
180 Queen's Gate,
London SW7 2BZ, United Kingdom}

\date{\today}

\begin{abstract}
Synchronization by exchange of pulses is a widespread phenomenon,
observed in flashing fireflies, applauding audiences and the neuronal
network of the brain. Hitherto the focus has been on integrate-and-fire
oscillators. Here we consider entirely analytic time evolution. Oscillators exchange narrow but finite
pulses.  For any non-zero time lag between the oscillators complete
synchronization occurs for any number of oscillators arranged in interaction networks whose adjacency
matrix fulfils some simple conditions. The time to synchronization decreases with increasing time lag.  
\end{abstract}



\maketitle

\section{Introdution}
The emergence of coherent structures in time and space though synchronization occurs across  the entire breadth of science: vibrating atoms, firing neurons, flashing fireflies, clapping audiences
etc. and has therefore been studied intensively from a mathematical viewpoint \cite{Strogatz:2003,Pikovsky:2001,Brechet:2000a}. 

Synchronization is often analyzed in models which explicitly favor phase
synchronization, e.g. in the seminal Kuramoto model \cite{Kuramoto:1975,Pikovsky:2001} and in diffusively coupled models (see e.g. \cite{Nishikawa_Motter:2010,Pereira:2011}). In these schemes the net-interaction between oscillators indeed vanishes in the synchronized state. 

However, in many cases, such as fireflies
\cite{Winfree:2001,Lewis:2008,Trimmer:2001}, cardiac cells, neuronal
system and applauding audiences \cite{Brechet:2000a,Brechet:2000b}  the
interaction between oscillators consists in the exchange of brief
pulses, which persist even when the system fully  synchronize. Since
Mirollo and Strogatz's influential 1990 paper
\cite{MirolloStrogatz:1990} such systems are often described by a set
of non-analytically evolving integrate-and-fire oscillators. Each
oscillator is described by a load variable, which is taken to have a
concave dependence on a monotonously increasing phase. When the load
reaches a certain threshold, relaxation occurs instantaneously and a
pulse is send to all connected oscillators. Receiving oscillators jump
discontinuously forward by a given amount. For such systems Mirollo and
Strogatz showed \cite{MirolloStrogatz:1990} that full synchronization
always occurs. Later Ernst, Pawelzik and Geisel
\cite{ErnstPawelzikGeisel:1995} demonstrated that for excitatory-only
couplings, synchronization depends on phase lag, whereas the presence of inhibitory couplings leads to full in-phase synchronization. 

The treatment of pulse oscillators in terms of non-analytic
integrate-and-fire oscillators is more a tradition than a necessity. In
the present paper we assume that each oscillator is represented by a
phase $\theta_i(t)$ whose time $t$ derivative is always equal to a
constant rate plus a sum of {\em smooth} but narrow pulses emitted by
surrounding oscillators coupled with strength $J$. Synchronization
(asymptotically vanishing phase difference) always occurs for this
system if pulses arrive with a non-zero time lag $\deltat$ for a very wide class of adjacencies, including the mean-field setting often considered in the literature.

Given the previous results for pulse oscillators
\cite{MirolloStrogatz:1990,ErnstPawelzikGeisel:1995} and to illuminate the more
detailed discussion below it is natural to begin our analysis of two
pulse exchanging phases by considering Dirac's delta pulses for the
interaction.
\begin{eqnarray}
\dot{\theta}_1(t) &=& \omega + J\sum_{n\in{\mathbb Z}}\delta(\theta_2(t-\delta t)-n))\nonumber\\
\dot{\theta}_2(t) &=& \omega + J\sum_{n\in{\mathbb Z}}\delta(\theta_1(t-\delta t)-n))
\label{two-delta}
\end{eqnarray}
Integrating  the time derivatives tells us that $\theta_1$ ``jumps''
each time $\theta_2$  passes through an integer value,
$\theta_1(t)\mapsto \theta_1(t)+J$, and vise versa for $\theta_2$. Let
$\theta_1(0)>\theta_2(0)$, it is straight forward to see,
\fref{Dirac_comb},  that the two phases are unable to synchronize though
in the case of a finite time delay they may leapfrog each other, as the
jump of one oscillator can make the other skip its. 

\begin{figure}[t]
\begin{center}
\includegraphics[width=8cm]{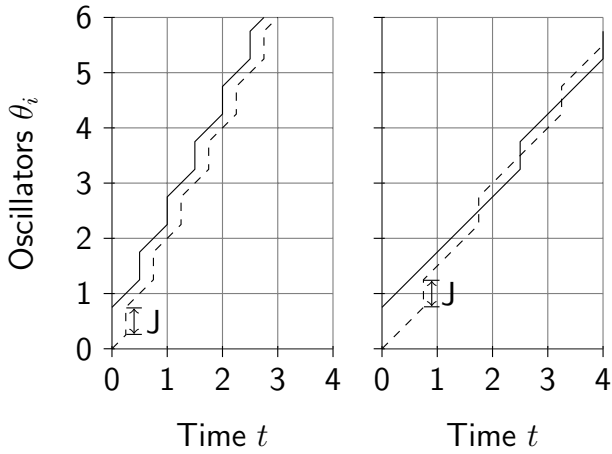}
\end{center}
\caption{\flabel{Dirac_comb}The time evolution of two oscilators (solid
and dashed lines) exchanging pulses according to Eq. \ref{two-delta}.
Left panel: $\delta t =0$, right panel: $\delta t=0.5$.}
\end{figure}

Obviously Dirac delta pulses are unrealistic. Pulses emitted by real
systems will have a finite width and a smooth time dependence
(\Eref{general_eom} below). The introduction of smooth pulses changes
the behavior in an essential way. As will be explained below
synchronization now takes place whenever a time lag is present, 
$\delta t>0$, and in this case complete synchronization occurs for all
smooth pulses.

\SectionHead{General model} 
We now consider $n$ coupled oscillators, each described by a single degree of freedom $\theta_i$, with $i=1,2,\dots,n$, and each with the same eigenfrequency $\omega$:
\begin{equation}
\dot{\theta}_i(t) = \omega + \sum_{j} J_{ij} \sigma(\theta_j(t-\deltat))
\elabel{general_eom}
\end{equation}
Oscillators are coupled through an adjancency matrix $J_{ij}$ and a feedback
function $\sigma(\theta)$ which has period $\anglePeriod$. It is only through  $\sigma(\theta)$ that periodicity is implemented:
$\sigma(\theta_i)$ describes the effect that the state of $\theta_i$ (say, the flashing of a firefly) has on any other oscillator. 
As opposed to other models often studied in synchronization, such as the Kuramoto model \cite{Pikovsky:2001}, the effect
of $\sigma$ does not disappear in the synchronized state.

We chose $\dot{\theta}_i>0$ at all times such that $\theta_i(t)$ are
monotonically increasing functions in time. This is achieved by
choosing $\sigma(\theta)>0$. In our numerical study below we use a comb of normalised Gaussians with period $\anglePeriod$ and width $w$, 
$\sigma(\theta)
= \sum_{n=-\infty}^\infty \exp{-(x+n\anglePeriod)^2/(2 w^2)} (2\pi
w^2)^{-1/2} 
= \anglePeriod^{-1} \vartheta_3(\pi x/\anglePeriod, \exp{-2 w^2 \pi^2/\anglePeriod^2)}$,
the Jacobi theta function.

In natural systems time delay is inevitable. We show that $\deltat>0$ is crucial for
synchronization.  We use this term in a strong sense: For any pair $i,
j$ of oscillators
$\lim_{t\to\infty} \theta_i(t)-\theta_j(t)=\mu_{ij} \anglePeriod$ with
$\mu_{ij}\in\Zset$, \ie the phase difference between any two oscillators
converges to an integer multiple of the period of $\sigma$, 
which implies $\lim_{t\to\infty} \dot{\theta}_i - \dot{\theta}_j=0$.
By inspection it is clear that for the synchronized state to exist indefinitely,
$\sum_{j} J_{ij}=\Jtilde$ needs to be independent of $i$, which means
that if synchronization takes place, the difference between any
$\theta_i(t)$ and the solution $\thetatilde(t)$ of
\begin{equation}
\dot{\thetatilde}(t) = \omega + \Jtilde \sigma(\thetatilde(t-\deltat))
\elabel{thetatilde_eom}
\end{equation}
with appropriate initial conditions vanishes asymptotically.
Provided $\Jtilde\ne0$, the eigenfrequency $\omega$ can be absorbed into $\sigma$, using
$\sigma(\theta)\to\sigma(\theta)+\omega/\Jtilde$.

\SectionHead{Simple two oscillator case} 
We now demonstrate that under very general conditions the system in Eq.
\Eref{general_eom} will synchronize in the long time limit. First we
consider the simple case of two oscillators, i.e. $n=2$ and
$J_{ij}=1-\delta_{ij}$.  By considering $\dot{\theta}_1/\dot{\theta}_2$,
it is easy to show that $\theta_1(t)-\theta_2(t)$ is periodic if
$\deltat=0$, \ie synchronization in the strong sense above does not
occur without time delay, rather, entrainment is inevitable. However, integrating the equation of motion \Eref{general_eom} numerically on the basis of a simple Euler method suggests differently. Better numerical schemes, such as the Runge Kutta \cite{PressETAL:1992} method, remove the spurious synchronization, which depends on the integration time step and therefore hints at the r{\^o}le of the time delay effectively implemented by the forward derivative used in the most na{\"i}ve Euler scheme.

We now analyze in detail the effect of a time delay by considering \Eref{general_eom} with $\deltat>0$. A linear stability analysis for small $\deltat$ and small deviations $\theta_i(t)-\thetatilde(t)$ reveals that any positive $\deltat$ leads
to a synchronized state.We present the calculation briefly in the following for  $n=2$ and
$J_{ij}=1-\delta_{ij}$. 

\begin{figure}[t]
\includegraphics*[width=\linewidth]{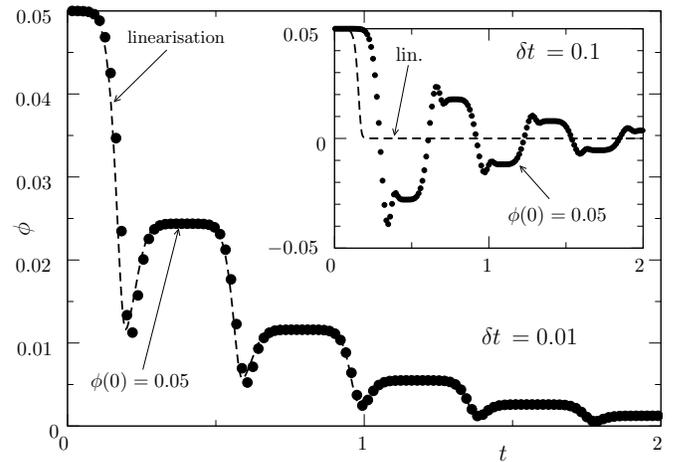}
\caption{\flabel{full_vs_lin}
Comparison of $\phi(t)=(\theta_1(t)-\theta_2(t))/2$ from a numerical 
integration of \Eref{general_eom} (filled circles) and the linear approximation
(dashed line) \Eref{phi_solution}, with $\phi(t)=0.05$ and $\deltat=0.01$.
Parameters are  $n=2$, $J_{ij}=1-\delta_{ij}$, $\omega=2$,
$\anglePeriod=1.01$ and $w=0.1$, so that $\psi\approx0.3934$.
The inset compares of full solution and linear approximation for $\phi(0)=0.05$ and
$\deltat=0.1$.  
}
\end{figure}

The equations of motion of  $\phi(t)=\half (\theta_1(t)-\theta_2(t))$ and $\thetabar=\half(\theta_1(t)+\theta_2(t))$ 
are
\begin{subeqnarray}{}
\dot{\phi}(t)&=&
\half (\sigma(\theta_2(t-\deltat))-\sigma(\theta_1(t-\deltat))) \\
\dot{\thetabar}(t)&=&
\half (\sigma(\theta_2(t-\deltat))+\sigma(\theta_1(t-\deltat))) +
\omega
\end{subeqnarray}
which we study to first order in $\phi$ and $\deltat$ and find (for details see \cite{PRE_version})   
\begin{equation}\elabel{phi_solution}
\phi(t) = \phi(0) \frac{\dot{\thetabar}(0)}{\dot{\thetabar}(t)} 
  \Exp{
    -2 \deltat \Jtilde^2
    \int_{\thetabar(0)}^{\thetabar(t)} \dint \theta 
    \frac{\sigma'^2(\theta)}{\omega + \Jtilde\sigma(\theta)}
  }
\end{equation}
with $t(\thetabar) = \int_{\thetabar(0)}^{\thetabar} \dint \theta
(\omega + \sigma(\theta))^{-1}$.
As the integrand is strictly positive, synchronization takes place in this approximation for all $\deltat>0$. \fref{full_vs_lin} shows that the linearized solution is a very good approximation of the full system
\Eref{general_eom}.

The characteristic time to synchronization is estimated in the following way. Define
\begin{subeqnarray}{}
\timePeriod &=& \int_0^\anglePeriod \dint \thetabar 
\frac{1}{\omega + \Jtilde\sigma(\thetabar)} \approx
\int_0^\anglePeriod \frac{\dint \thetabar}{\dot{\thetabar}}
\\
\syncPeriod &=& \int_0^\anglePeriod \dint \thetabar
\frac{\sigma'^2(\thetabar)}{\omega + \Jtilde\sigma(\thetabar)}
\end{subeqnarray}
where $\timePeriod$, to leading order, is the time for
$\sigma(\thetabar(t))$ to go through one period, \ie
$\thetabar(t+\timePeriod)\approx \thetabar(t)+\anglePeriod$.  
 $\syncPeriod$ corresponds to the integral in the exponent of
\Eref{phi_solution} for $\thetabar(t)=\thetabar(0)+\anglePeriod$. 
As the integrand is periodic we estimate
\begin{equation}
    \int_{\thetabar(0)}^{\thetabar(t)} \dint \theta 
    \frac{\sigma'^2(\theta)}{\omega + \Jtilde\sigma(\theta)}
    \approx
    \frac{t}{\timePeriod} \syncPeriod
\end{equation}
and rewrite \Eref{phi_solution}
$\phi(t) = \phi(0)
(\dot{\thetabar}(0)/\dot{\thetabar}(t)) \exp{-t/\syncTime}$,
with the characteristic synchronization time 
\begin{equation}
\syncTime \simeq \frac{\timePeriod}{2 \Jtilde^2 \deltat \syncPeriod} \ ,
\elabel{syncTime}
\end{equation}
inversely proportional to the time delay $\deltat$.

\SectionHead{Network of oscillators}
The above picture can be  extended to arbitrary coupling matrices
$J_{ij}$, or a (weighted) network adjacency matrix. The only constraint
is $\sum_{j} J_{ij}=\Jtilde$ independent from $i$, similar to a Markov
matrix. Motivated by the observation that the two parameters used above,
$\thetabar$ and $\phi$, are based on the eigenvectors of the  matrix
$J_{ij}=1-\delta_{ij}$ studied for $n=2$, we consider the time evolution
of $\braket{i}{\phi(t)}$, \ie of ``normal modes'', where $\bra{i}$ is
the $i$th left eigenvector of $J$, which, we assume for simplicity, has
$n$ linearly independent eigenvectors. For simplicity, we normalize
$\braket{i}{j}=\delta_{ij}$. The matrix $J$ is not necessarily symmetric
so generally $\left(\bra{i}\right)^{\dagger}\ne\ket{i}$. Due to the Markov property, there is a pair of left and right eigenvectors with eigenvalue $\Jtilde$, which in the following is denoted by $\bra{1}$ and
$
\ket{1}=\sum_i^n \ket{e_i}
$
respectively, where $\ket{e_i}$ denotes the canonical basis of the
$\Rset^n$.  

The state of the entire system is written in vector form as
$\ket{\theta(t)} = \sum_i^n \ket{e_i} \theta_i(t)$. The column vector $\ket{\phi(t)}$ is the deviation 
$\ket{\phi(t)} = \ket{\theta(t)} - \thetabar(t) \ket{1}$ of $\ket{\theta(t)}$ from
$\thetabar(t)=\braket{1}{\theta(t)} $ anticipating that $\thetabar(t)$ represents the asymptotically
synchronised state. Following the procedure above, one finds  
\begin{multline}
\braket{i}{\phi(t)} = A_i
\left(
\frac{T(\thetabar(t))}{T_0}
\right)^{\frac{\lambda_i}{\Jtilde}} \\
\times \Exp{\lambda_i(\Jtilde - \lambda_i)\delta t
\int_{\thetabar(0)}^{\thetabar(t)} \frac{\sigma'^2(\theta')}{T(\theta')}
\dint{\theta'} }
\elabel{general_linear_solution}
\end{multline}
where 
$T_0=T(\thetabar(0))$ and
$T(\thetabar)= \omega + \Jtilde \left( \sigma(\thetabar) - \deltat
\dot{\thetabar}(\thetabar)\sigma'(\thetabar)\right) =
\dot{\thetabar}(\thetabar) + \OC(\deltat^2)$.

The amplitudes $A_i$ are determined by the
initial projections $\braket{i}{\phi(0)} = A_i$. 
\Eref{general_linear_solution} also applies to $i=1$, 
yet $\braket{1}{\phi(t)}=0$ by construction so that
$A_1=0$.
The special case $\Jtilde=0$ (so that $\dot{\thetabar}=\omega$ to
linear order) coincides with the limit $\Jtilde\to0$,
where
\begin{equation}
\lim_{\Jtilde\to0}
\left(\frac{T(\thetabar(t))}{T_0}\right)^{\frac{\lambda_i}{\Jtilde}}=
\exp{\lambda_i \sigma(\thetabar)/\omega - \deltat \lambda_i
\sigma'(\thetabar)}
\end{equation}
to leading order, assuming $T_0=\omega$ for simplicity.

Since $T(\thetabar)$ is periodic, the long-term behaviour of $\braket{i}{\phi(t)}$
depends crucially on the sign of the real part of $\lambda_i(\Jtilde - \lambda_i)$. If it
is negative, the projection has an approximate synchronization time
\begin{equation}
\syncTime_i = \frac{\int_0^{\anglePeriod} \dint{\thetabar} \frac{1}{\omega +
\Jtilde \sigma(\thetabar)}}{-\RealPart{\lambda_i(\Jtilde - \lambda_i)} \deltat 
\int_0^\anglePeriod \dint \thetabar
\frac{\sigma'^2(\thetabar)}{\omega + \Jtilde \sigma(\thetabar)}} \ ,
\end{equation}
corresponding to \Eref{syncTime}. Here $\RealPart{\cdot}$ denotes the
real part. The usual mean-field setup $J_{ij}=a (1-\delta_{ij})$ has one eigenvalue
$\Jtilde=(n-1)a$ and $n-1$ eigenvalues $\lambda_i=-a$, so that
$\lambda_i(\Jtilde - \lambda_i)=-n a^2$ has a negative real part
provided $a^2$ has a positive one, \ie in particular for all real $a$.
The mean-field theory thus always synchronises, as does the special case
of a lattice Laplacian.


\begin{figure}[t]
\includegraphics*[width=\linewidth]{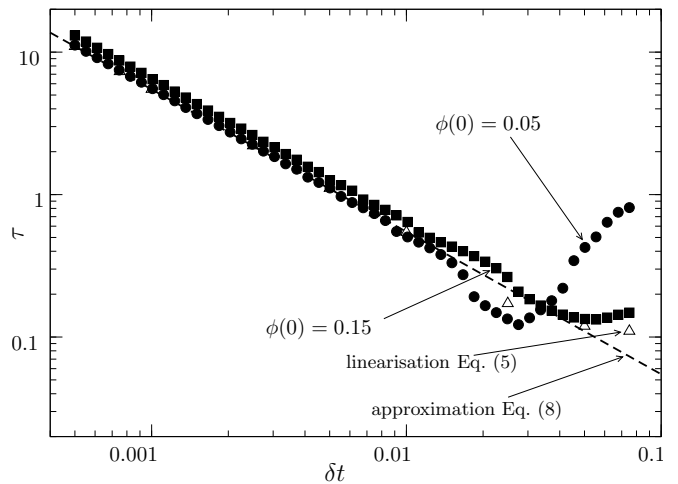}
\caption{\flabel{sync_time} 
Plot of the synchronization time estimated from
the window averaged phase difference $\theta_1(t)-\theta_2(t)$ (average
taken over a time period $t^*$ so that
$\thetabar(t)=\thetabar(t-t^*)-\xi$). The filled symbols refer to
results based on \Eref{general_eom}, the empty triangles to  
\Eref{phi_solution} and the line to   \Eref{syncTime}. 
Parameters as in \fref{full_vs_lin}.
\vspace{-.5cm}}
\end{figure}

The perturbative result \Eref{general_linear_solution} can be compared to the numerical
integration of the system. We used a fourth order Runge-Kutta integration scheme
\cite{PressETAL:1992,PRE_version} and show in \fref{sync_time} that
the derived synchronization time compares very well (for time delays up
to 5\% to 10\% of the  synchronized period) to that of the linearised result, \Eref{phi_solution} and to the estimate \Eref{syncTime}.

\SectionHead{Mechanism}
How is synchronization achieved? \fref{mechanism_theta}
shows $\sigma(\theta_i(t))$ and $\dot{\theta_i}(t)$ as a function of $t$ for $n=2$. Synchronization occurs
because $\theta_2$ experiences a greater increase in speed by
$\sigma(\theta_1(t-\deltat))$ than $\theta_1$ does by
$\sigma(\theta_2(t-\deltat))$.  This asymmetry comes about because
$\theta_2$ is relatively fast itself when $\sigma(\theta_2(t))$
goes through its maximum and $\theta_1$ is relatively slow when
$\sigma(\theta_1(t))$ goes through its maximum. As a result
the maximum $\sigma(\theta_1(t))$ is broadened as a function of
time, and $\sigma(\theta_2(t))$ is narrowed (this effect is minute and
thus not visible in \fref{mechanism_theta}). Therefore, 
the maximum of $\sigma(\theta_1(t-\delta))$  enters into
$\dot{\theta_2}$ for a longer time period than $\sigma(\theta_2(t-\delta))$
enters into $\dot{\theta_1}$, leading to a speedup of $\theta_2$
relative to $\theta_1$. 
In summary, synchronization is result of 
oscillator $i$ being slow or fast when going through the maximum of the
function $\sigma(\theta_i)$.
What r{\^o}le has the time delay in this?
The time delay ensures that the trailing oscillator $\theta_2$ 
receives a boost at a time when $\sigma(\theta_2(t))$ goes
through a maximum, while the leading $\theta_1$ receives its boost at
a time when $\sigma(\theta_1(t))$ goes through a local
minimum. Without the time delay, the effect of the speed-up and the
slowdown would indeed be perfectly symmetric.

We notice that the mechanism underlying the synchronization supported by 
Eq. \eref{general_eom} is a kind of Doppler effect that makes the received pulse change its duration when the sending oscillator changes its speed.

\begin{figure}
\includegraphics[width=0.95\linewidth]{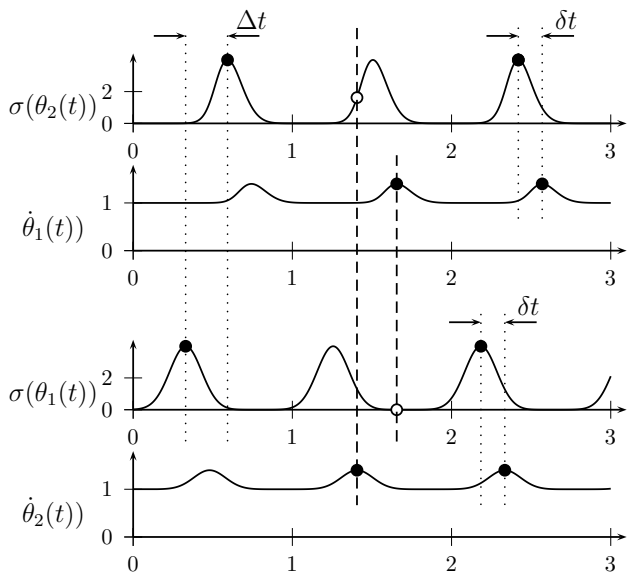}
\caption{\flabel{mechanism_theta}
We see that
$\theta_1$ is ahead of $\theta_2$, with the maximum of
$\sigma(\theta_2(t))$ displaced by $\Delta t$ (as indicated) to the right relative to
that of $\sigma(\theta_1(t))$, as initialised. At a given time,
$\theta_2$ still has to
pass through the maximum of $\sigma(\theta_2(t))$ when $\theta_1$
already has. The phase speed $\dot{\theta}_{1,2}(t)$ is essentially
$\sigma(\theta_{2,1}(t))$ shifted by $\deltat$ to the right, as
indicated by the dotted lines. As a
result, the maximum of $\dot{\theta}_2$ nearly aligns with the maximum of
$\sigma(\theta_2(t))$, \ie $\theta_2$ is fast when $\sigma(\theta_2(t))$
goes through the maximum (dashed line), thereby narrowing it. In turn, 
$\theta_1$ passes very quickly through a low value of
$\sigma(\theta_1(t))$, relatively broadening in turn the maximum of
$\sigma(\theta_1(t))$.
\vspace{-.5cm}}
\end{figure}

Equation \Eref{general_eom}  provides a remarkably simple
mechanism for synchronization. Because oscillators lagging behind by a
certain amount catch up in every period of $\sigma(\thetabar)$
by an amount of phase difference proportional to the phase
difference at the beginning of the period, the model can immediately be
extended to one with different eigenfrequencies $\omega_i$ of oscillators or
some variation in $\sum_j J_{ij}$ with $i$ or of $\sigma$ and even
$\deltat$. An analysis of such extensions follows the derivation above, see \cite{PRE_version}. 
It will generally lead to entrainment.

Synchronisation by time delay is a viable explanation for natural
synchronisation phenomena whenever oscilators respond to the duration of
the pulse received. Fireflies are known to be able to change the pulse
duration and female fireflies are sensitive to that \cite{Lewis:2008}.
The exact way a clapping audience reaches synchrony
\cite{Brechet:2000a,Brechet:2000b} can be analyzed sufficiently
accurately to establish whether people change the duration of the
individual clap \cite{PeltolaETAL:2007} in the process of reaching
synchrony.


\end{document}